\documentclass[structabstract]{aa}  

\usepackage{graphicx}
\usepackage{txfonts}
\usepackage{natbib}
\usepackage{rotating}
\newcommand{\kms}{km\ s$^{-1}$}

\newcommand{\xmm}{{\sc XMM}\emph{-Newton}}

\newcommand{\hei}{He\,{\sc i}}
\newcommand{\heii}{He\,{\sc ii}}
\newcommand{\ciii}{C\,{\sc iii}}
\newcommand{\niii}{N\,{\sc iii}}
\newcommand{\civ}{C\,{\sc iv}}
\newcommand{\ha}{H$\alpha$}
\newcommand{\hb}{H$\beta$}
\newcommand{\hg}{H$\gamma$}
\newcommand{\lxlb}{$L_X/L_{BOL}$}
\newcommand{\loglxlb}{$\log[L_X/L_{BOL}]$}

\begin{document}
  \title{New findings on the prototypical Of?p stars\thanks{Based on observations collected at the Haute-Provence Observatory, at the La Silla and San Pedro M\'artir  Observatories, and with XMM-Newton, an ESA Science Mission with instruments and contributions directly funded by ESA Member States and the USA (NASA).}}

   \author{Ya\"el Naz\'e
          \inst{1}\fnmsep\thanks{Research Associate FRS-FNRS} \and Asif ud-Doula\inst{2} \and Maxime Spano\inst{3} \and Gregor Rauw\inst{1}\fnmsep$^{\star\star}$ \and Micha\"el De Becker\inst{1,4} \and Nolan R. Walborn\inst{5}
          }

   \institute{GAPHE, D\'epartement AGO, Universit\'e de Li\`ege, All\'ee du 6 Ao\^ut 17, Bat. B5C, B4000-Li\`ege, Belgium\\
              \email{naze@astro.ulg.ac.be}
   \and Penn State Worthington Scranton, 120 Ridge View Drive, Dunmore, PA 18512, USA
   \and Observatoire de Gen\`eve, Universit\'e de Gen\`eve, 51 Chemin des Maillettes, CH1290-Sauverny, Switzerland
   \and Observatoire de Haute-Provence, F04870-St Michel l'Observatoire, France 
   \and Space Telescope Science Institute\thanks{Operated by the Association of Universities for Research in Astronomy, Inc., under NASA contract NAS5-26555.}, 3700 San Martin Drive, Baltimore, MD 21218, USA
             }

%   \date{}

% \abstract{}{}{}{}{} 
% 5 {} token are mandatory
 
  \abstract
   {} 
   {In recent years several in-depth investigations of the three Galactic Of?p stars were undertaken. These multiwavelength studies revealed the peculiar properties of these objects (in the X-rays as well as in the optical): magnetic fields, periodic line profile variations, recurrent photometric changes. However, many questions remain unsolved.}
   {To clarify some of the properties of the Of?p stars, we have continued their monitoring. A new \xmm\ observation and two new optical datasets were obtained. }
   {Additional information for the prototypical Of?p trio has been found. \\ HD\,108 has now reached its quiescent, minimum-emission state, for the first time in 50--60yrs. \\ The \'echelle spectra of HD\,148937 confirm the presence of the 7d variations in the Balmer lines and reveal similar periodic variations (though of lower amplitudes) in the \hei\,$\lambda$\,5876 and \heii\,$\lambda$\,4686 lines, underlining its similarities with the other two prototypical Of?p stars. \\ The new \xmm\ observation of HD\,191612 was taken at the same phase in the line modulation cycle but at a different orbital phase as previous data. It clearly shows that the X-ray emission of HD\,191612 is modulated by the 538d period and not the orbital period of 1542d - it is thus not of colliding-wind origin and the phenomenon responsible for the optical changes appears also at work in the high-energy domain. There are however problems: our MHD simulations of the wind magnetic confinement predict both a harder X-ray flux of a much larger strength than what is observed (the modeled DEM peaks at 30-40\,MK, whereas the observed one peaks at 2\,MK) and narrow lines (hot gas moving with velocities of 100--200\kms, whereras the observed FWHM is $\sim$2000\kms). }
   {}


   \maketitle
%
%________________________________________________________________

\section{Introduction}
The presence of \ciii\,$\lambda$\,4650\AA\ in emission with a strength comparable to the neighbouring \niii\ lines is not a usual feature of O-star spectra. In the past, stars displaying this feature were classified by \citet{wal72} in a separate category, dubbed Of?p. In recent years, this category has attracted a lot of attention, as exemplified by numerous publications \citep{wal03,wal04,don06,how07,hub08,naz01,naz04,naz07,naz08}. These studies unveiled the peculiar properties of these stars \citep[for a summary see][]{naz08b} and led to a modification of the definition of these stars. The Of?p phenomenon now covers stars presenting recurrent spectral variations (in the Balmer, \hei, \ciii, Si\,{\sc iii} lines), strong \ciii\,$\lambda$\,4650\AA\ in emission (at least at some phases), narrow P Cygni or emission components in the Balmer hydrogen lines and \hei\ lines (at least at some phases), and UV wind lines weaker than for typical Of supergiants \citep{naz08,wal10}. Stars that appear otherwise normal but display strong \ciii\,$\lambda$\,4650 emission are now called ``Ofc'' stars \citep{wal10}.

The three prototypical Of?p stars are HD\,108, HD\,148937, and HD\,191612. They are the most studied objects of that category, though some open questions still remain unanswered.

HD\,108 displays correlated photometric, Balmer, and \hei\ line profile variations: the star is brighter when the emission lines are stronger \citep{naz01,naz08b}. The variations appear recurrent, with a probable time scale of about 55 years \citep{naz06}. The minimum emission state was not yet reached in 2004 \citep{naz04}, preventing a precise evaluation of the recurrence time. Recently, an additional observational result enriched the picture of HD\,108: the detection of a strong magnetic field \citep[observed strength is 100--150\,G, corresponding to a potential dipolar field of 0.5--2\,kG][]{mar10}.

The behaviour of HD\,191612 appears very similar to that of HD\,108, but with a shorter, well-defined period of 537.6d \citep[and references therein]{how07}. A strong magnetic field ($-$220$\pm$38\,G) was detected for this star by \citet{don06}, who further proposed an oblique magnetic rotator model for the star. Following these authors, the long period would result from magnetic braking of the stellar rotation \citep[for an analysis of such effects see][]{udd09}. Studies are now under way to check the exact geometry of the magnetic field (G. Wade, private communication). Finally, it is interesting to note that this star has a lower-mass companion in a 1542d orbit \citep{how07}.

HD\,148937 does not display the large photometric changes and line profile variations unlike the two other objects of this study. Instead, changes of limited amplitude (peak's amplitude varying by a few percents vs several 100\% for the other prototypical Of?p stars) were observed for the \ha\ line, with a possible period of about 7d, during a long-term monitoring campaign \citep{naz08}, but the temporal sampling of the optical dataset was not optimized for finding such a short period and the resultant value could therefore be subject to large errors. In addition, no change was detected in the low-resolution data for other spectral lines, which differs from what is observed in HD\,108 and HD\,191612. Finally, a tentative detection of a magnetic field was reported for this star by \citet[; $B_z=-276\pm88$\,G]{hub08}, but a spectro-polarimetric monitoring throughout the 7d cycle is clearly needed to ascertain the strength and geometry of the field. 

In addition to dedicated optical campaigns, several \xmm\ observations of these three objects were obtained. The derived properties are remarkably similar: overluminosities with respect to the `canonical' \lxlb $=10^{-7}$ relation, soft spectra, broad X-ray lines \citep{naz04,naz07,naz08}. Monitoring HD\,191612 further revealed small variations of the X-ray flux. However, as all data were taken during the same orbital cycle and the same 538d cycle, the exact cause for these changes remained unclear.

Though the published studies have led to the characterization of many aspects of these Of?p stars, several questions still remain unanswered - usually, these questions are different for the different objects (e.g. on the variations of the X-rays of HD\,191612, on the evolution of the optical spectra of HD\,108, or on the variation timescale of HD\,148937), due to the inhomogeneous datasets reported up to now. In the past few years, we have continued our monitoring of these three peculiar objects, with the objective of characterizing and better understanding their properties. This paper reports the results of this campaign. Section 2 presents the datasets and Section 3 the results; Section 4 summarizes the new information.

\begin{figure*}
\begin{center}
\includegraphics[width=15cm,bb=24 470 575 690, clip]{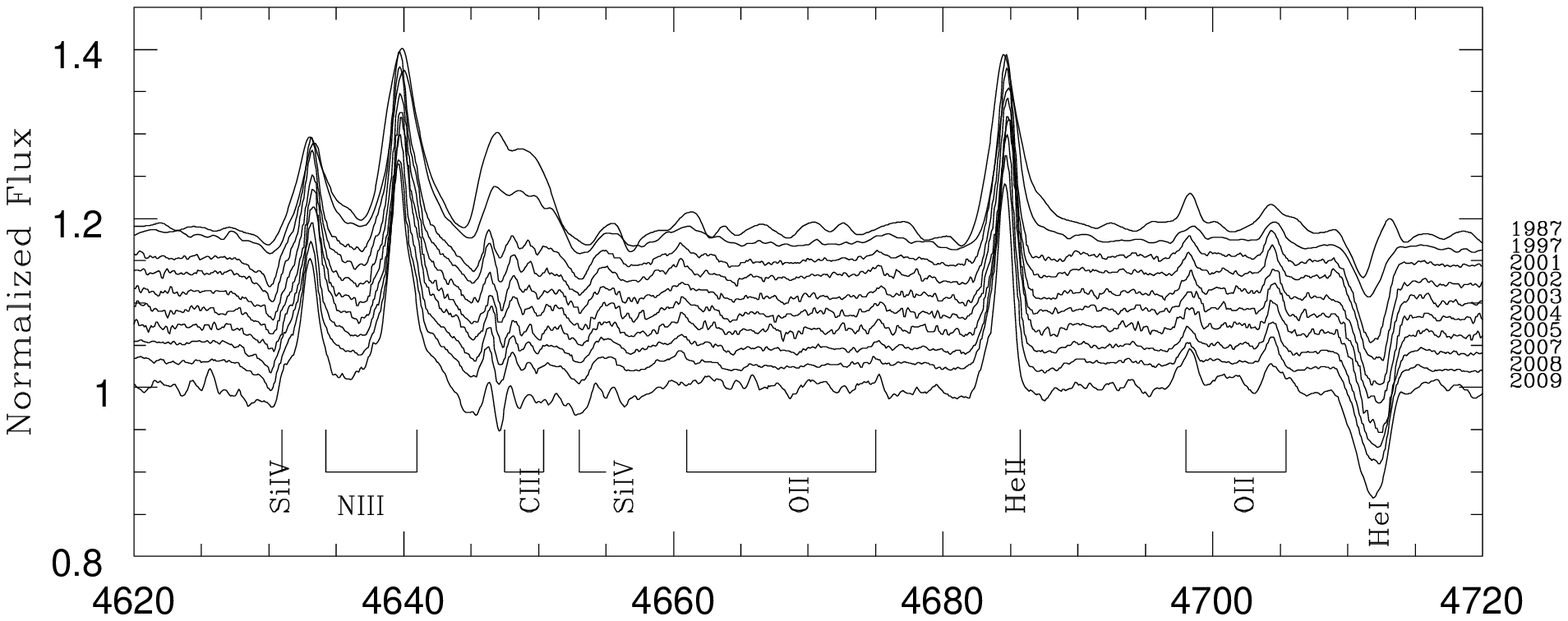}
\includegraphics[width=15cm, bb=25 180 575 430]{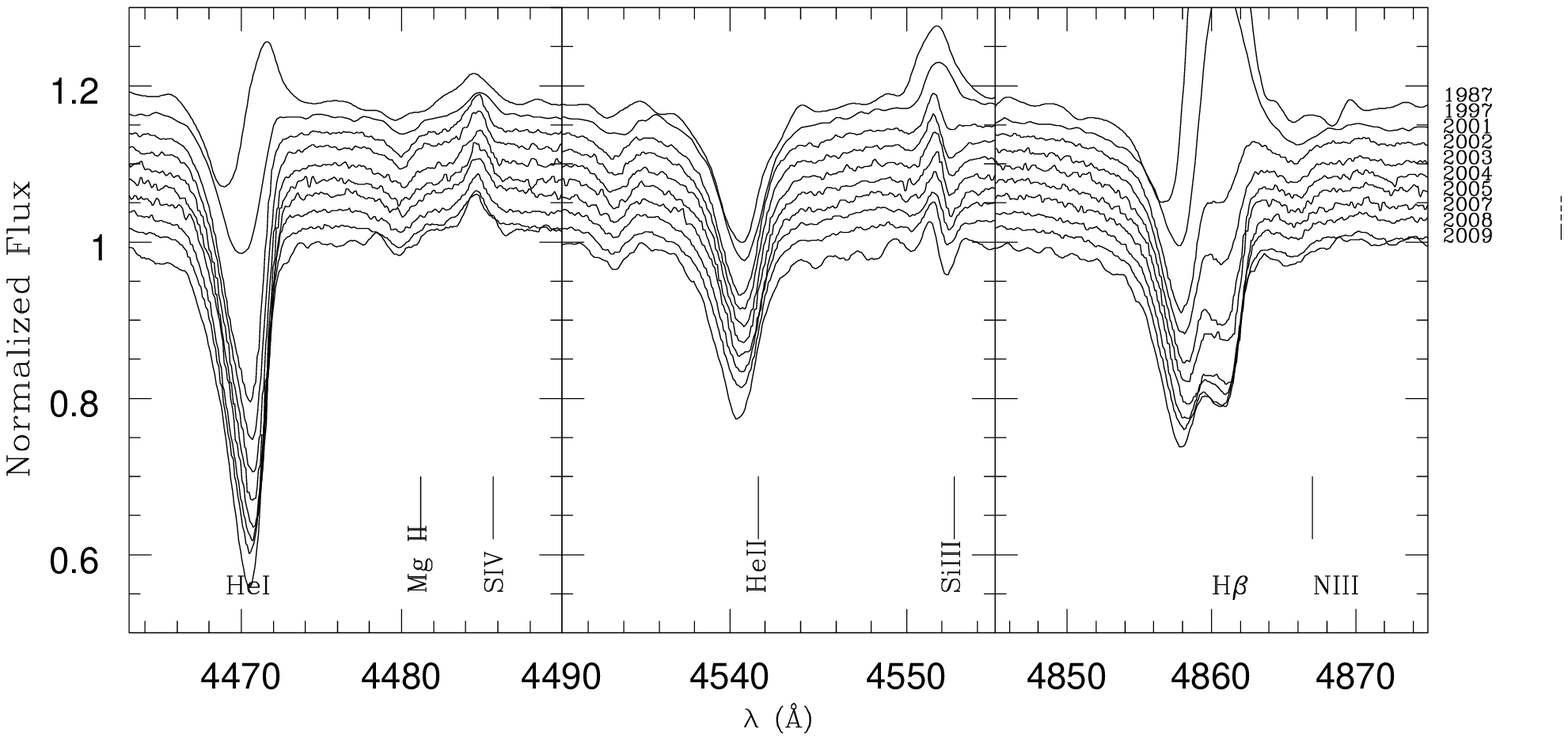}
\includegraphics[width=15cm, bb=18 140 550 430, clip]{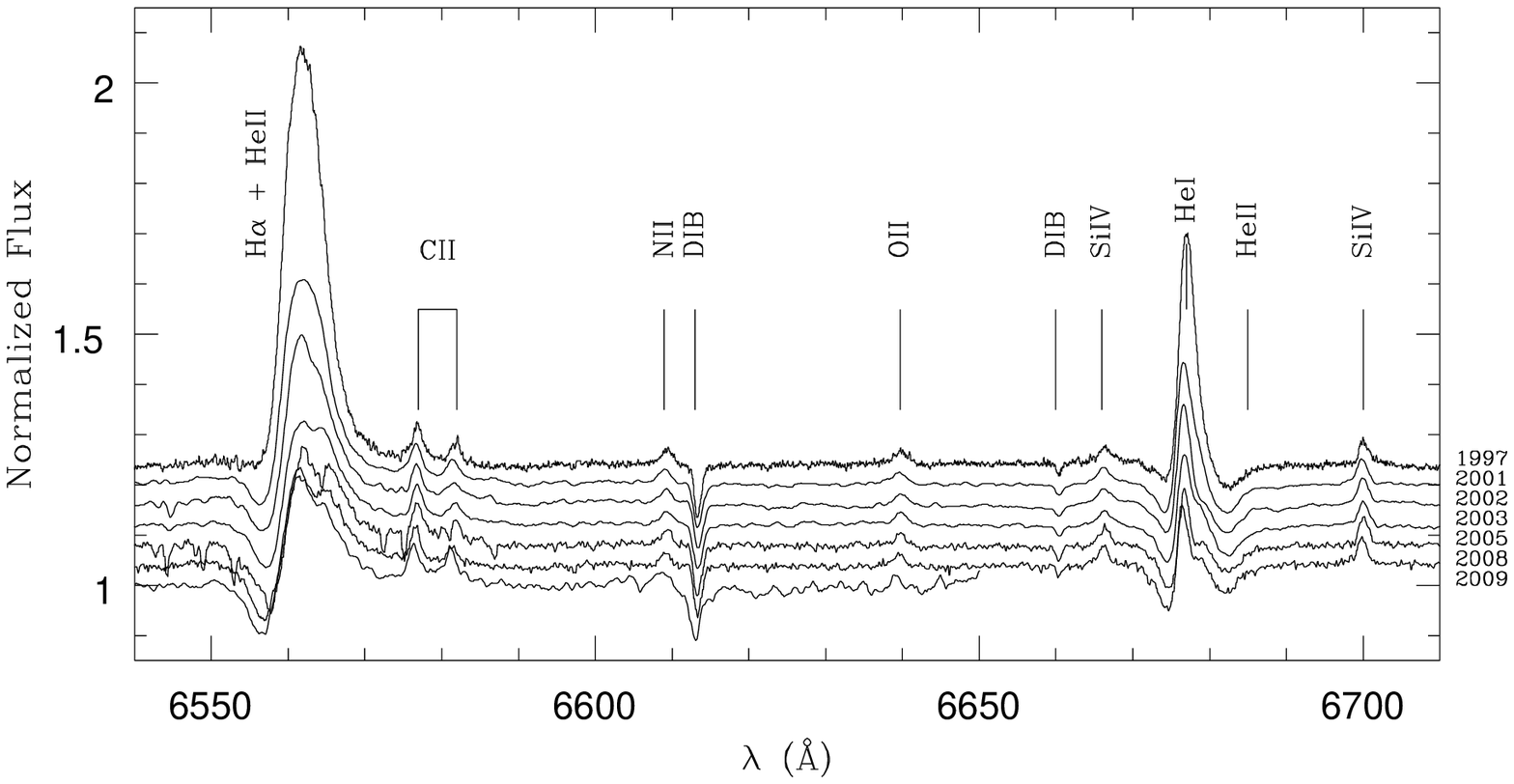}
\caption{\label{108prof} Evolution of the line profiles in recent years for HD\,108.}
%The dashed lines show the data from 1987 and 1997, the dotted lines the data between 2001 and 2004, the solid line the data since 2005 (black - 2005, red - 2007, 2008 ; the colored version of the figure is available on-line).} 
\end{center}
\end{figure*}

\section{Observations and data reduction}
\subsection{Aur\'elie data}
Since 2004, when optical spectra of HD\,108 were last published \citep{naz04}, we have continued the monitoring of the star with the Aur\'elie spectrograph mounted on the 1.52m telescope of the Haute-Provence Observatory (OHP, France). The detector was a thin back-illuminated CCD composed of 2048$\times$1024 pixels of size $13.5\,\mu$m$^2$. We used the grating \# 3, reaching a resolving power between 8000 and 10\,000 in the blue and red part of the spectrum, respectively. With this instrumental setup, we obtained spectra in the 4450--4900\AA, 5480--5920\AA, and 6340-6780\AA\ ranges. Exposure times were typically of 20--40 min, resulting in a signal-to-noise ratio of about 200 on average. The data were reduced in a standard way using MIDAS. The wavelength calibration was done using ThAr exposures taken immediately before/after the stellar spectrum, and the spectra were normalized using low-degree polynomials.

An additional \'echelle spectrum obtained in June 2009 at San Pedro M\'artir with the Espresso spectrograph was also made available to us by P. Eenens and L. Mahy. This dataset has been added to the analysis and discussion.

\subsection{Coralie \'echelle spectra}
Twenty \'echelle spectra of HD\,148937 were obtained in 2008 May 8--21 with the 1.2\,m Euler Swiss telescope at La Silla (Chile), equipped with the Coralie spectrograph and an EEV 2k$\times$2k CCD with pixel size $15\,\mu$m$^2$. The Coralie instrument is an improved version of the \'Elodie spectrograph \citep{bar96}. These observations covered the spectral range 3850--6890\AA\ with a resolving power of 55\,000. During the first week, two spectra per night were taken; during the second week, one spectrum per night was taken. The integration times were fixed to 30min (except for the first two exposures, which were each 20 min long) and the typical S/N ratio at 5000\AA\ was 150. The data were first reduced with the Coralie pipeline and then normalized using low-degree polynomials. A telluric correction was applied for \ha\ and \hei\,$\lambda$\,5876 using the telluric spectrum of \citet{hinkle00} and the task $telluric$ within IRAF.

\subsection{\xmm\ observation}
HD\,191612 was observed anew by \xmm\ on 2008 April 2 (Rev. 1523, ObsID 050068201, PI Naz\'e), for a total exposure time of 24\,ks. The instrumental configuration, processing, and extraction regions were identical to the first four observations of the source \citep{naz07}. We processed the data with the Science Analysis System (SAS) software, version~8.0; further analysis was performed using the FTOOLS tasks and the XSPEC software v 11.2.0. No high background episode (due to soft protons) was detected at high-energies during this exposure. 

In addition to EPIC data, \xmm\ provides high-resolution spectra thanks to the RGS instruments. The RGS data were reduced with the SAS and then combined with the previous observations following the procedure outlined in \citet{naz07}. However, the improvement in signal-to-noise of the combined spectrum is not important as all observations are of similar duration. We therefore have no new results to report on the high-resolution spectrum of HD\,191612, a quantitative leap in quality awaiting the advent of more sensitive observatories. 

\section{New results on Of?p stars}
\subsection{HD\,108}

The spectrum of HD\,108 has been followed intensively since 1986 \citep{naz01,naz04}. While some (purely photospheric) lines such as \heii\,$\lambda$\,4542 remain constant, the strength of Balmer, \hei\ (e.g. at $\lambda\lambda$\,4471,4713,6678), Si\,{\sc iii} (e.g. at $\lambda$\,4552) and \ciii\,$\lambda\lambda$\,4650 lines have decreased with time, with profiles changing from pure emissions or P Cygni profiles to strong absorptions or greatly reduced emissions, as illustrated by Fig. \ref{108prof}. These lines thus appear as a blend of a constant photospheric absorption and a varying emission component. A similar change was observed about 55 years ago \citep{naz01,naz06} but due to the lack of detailed monitoring in the past - often, only a global idea of the profile (emission vs absorption) was recorded - the timescale of the variations could not be precisely assessed and some even questioned the recurrence of the phenomenon. It is therefore of utmost importance to see how the lines evolve and if quiescence is reached. 

\begin{figure}
\begin{center}
\includegraphics[width=9cm]{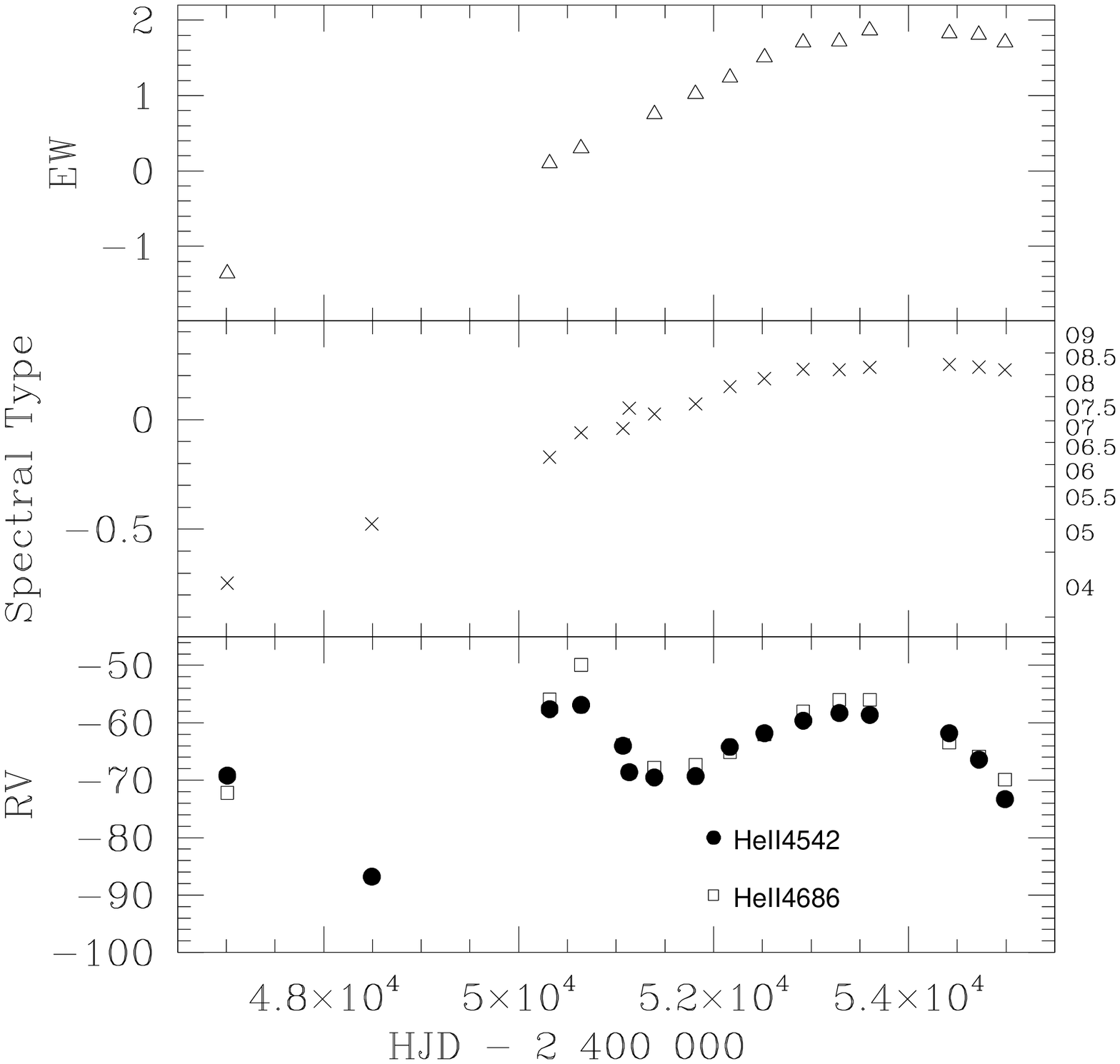}
\caption{\label{108hb} Evolution as a function of time of a few quantities related to HD\,108. Upper panel: EW (in \AA, in the range 4845--4870\AA) of \hb. Middle panel: spectral type criterion $\log \left( \frac{EW_{HeI4471}}{EW_{HeII4542}}\right)$. Lower panel: RVs (in \kms) of \heii\,$\lambda\lambda$\,4542,4686.} 
\end{center}
\end{figure}

Though the emission component had nearly disappeared these last years \citep{naz04}, the emissions continued to decline until 2004. Now, HD\,108 has finally reached its quiescent state since the data taken between 2005 and 2009 are very similar (Figs. \ref{108prof} and \ref{108hb}). Based on past behaviour \citep{naz01}, it is expected that the emission will strengthen again in the future, as is seen in HD\,191612. We note however that, for HD\,191612, the quiescent state occurs during one-third of the cycle, which would correspond to $\sim$18\,yrs for HD\,108. Assuming that the variations are due to an oblique magnetic rotator configuration, this quiescence timescale would however depend on the exact geometry of the system, which may be different between the two objects despite their obvious similarities. from literature, it is only known that \hb\ remained in absorption during about 10 years, while \hei\,$\lambda$\,4471 did so during several decades \citep{naz01}. However, the exact evolution was not recorded and it cannot be concluded from the older data how long the lines remained constant, in a quiescent state.
%After several years which saw the emission components of the line profiles declining \citep{naz01,naz04,naz06}, HD\,108 has finally reached its quiescent state. Indeed, the data taken between 2005 and 2009 are very similar (Figs. \ref{108prof} and \ref{108hb}). One question remains open: will the emissions of HD\,108 remain low for some time, as seen in HD\,191612 \citep[see e.g. ][]{how07}, or will they increase just after reaching the minimum? 

\subsection{HD\,148937}

Previously, a long-term monitoring (3 yrs, with monthly observations) provided a large set of low-resolution spectra. These data revealed the variability of 3 lines: \ha, \hb, and \heii\,$\lambda$\,4686 \citep{naz08}. However, this variability was of low amplitude: the EWs measured for \ha, the most variable line, display a dispersion which, though it was 10 times larger than that observed for the neighbouring narrow Diffuse Interstellar Band (DIB), correspond to a change of only 20\% in EW (it also corresponds to a variation in the peak's amplitude of only 7\% between the two extreme profiles), whereas the other prototypical Of?p stars have \ha\ lines varying from absorption to emission, with EWs and peak's amplitudes varying by $>>$100\%. Moreover, the \hei\ and \ciii\ lines seemed relatively constant, a behaviour different from that of HD\,108 and HD\,191612  - though subtle changes may have remained hidden in such low-resolution data. A period search further revealed a period of 7.031$\pm$0.003d when studying closely the line profile variations of \ha. However, the data sampling, aimed at studying monthly-to-yearly variations, was not adequate for detecting such a short period, which thus required confirmation with a more intense temporal sampling. A new, short-term monitoring was therefore undertaken, yielding 20 high-resolution spectra over 2 weeks.

\begin{figure*}
\begin{center}
\includegraphics[width=9cm]{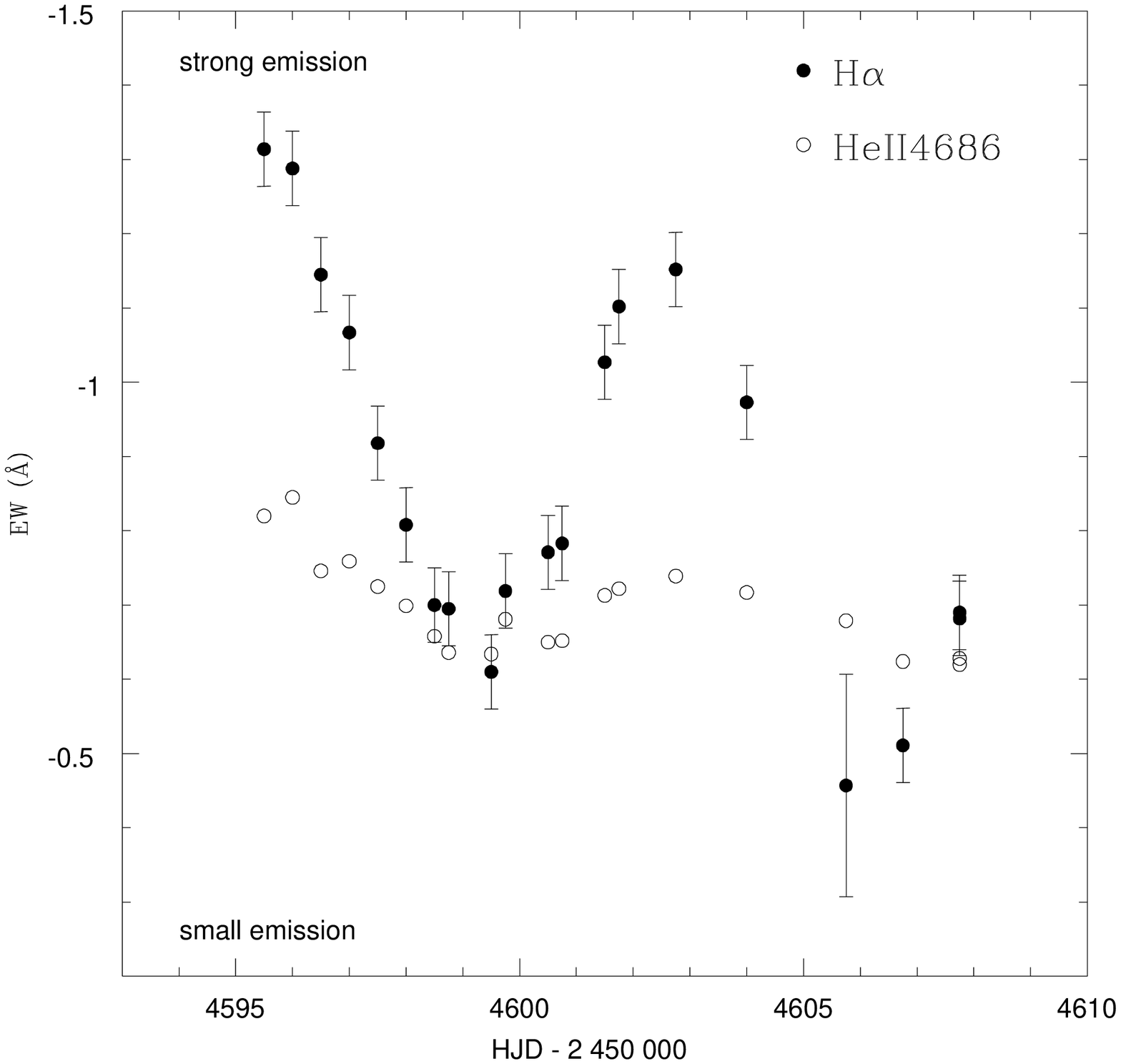}
\includegraphics[width=9cm]{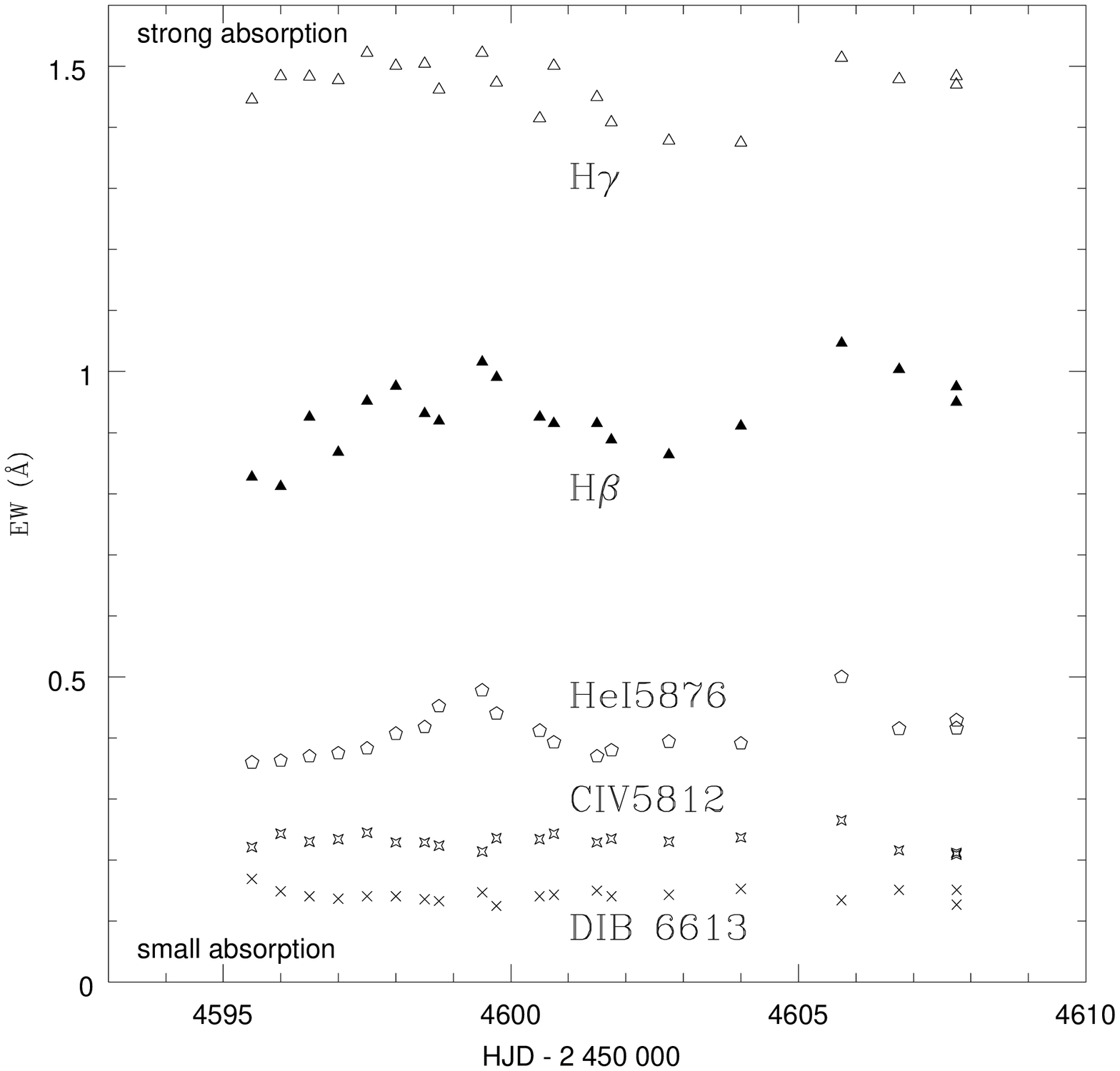}
\caption{\label{148937rvew} Evolution with time of the EWs (expressed in \AA) for some lines of HD\,148937. Left: The strongest emission lines (Balmer line of \ha\ and \heii\,$\lambda$\,4686) ; the emission strength increases from bottom to top. Right: Some absorption lines ; the absorption increases from bottom to top. The \hei\,$\lambda$\,5876 and \ha\ data were corrected for the contamination by telluric lines.} 
\end{center}
\end{figure*}

For consistency, the radial velocities and equivalent widths (EWs) were estimated as in \citet{naz08}; they are given in Table \ref{tab:148937}. While the \civ\ and DIBs display constant EWs, the Balmer, \heii\,$\lambda$\,4686, and \hei\,$\lambda$\,5876 lines show a larger dispersion. When plotted against time (Fig. \ref{148937rvew}), obvious modulations are detected. \ha\ and \heii\,$\lambda$\,4686 lines display maximum emission when \hb, \hg, and \hei\,$\lambda$\,5876 present a minimum absorption.

\begin{figure*}
\begin{center}
\includegraphics[width=10.5cm]{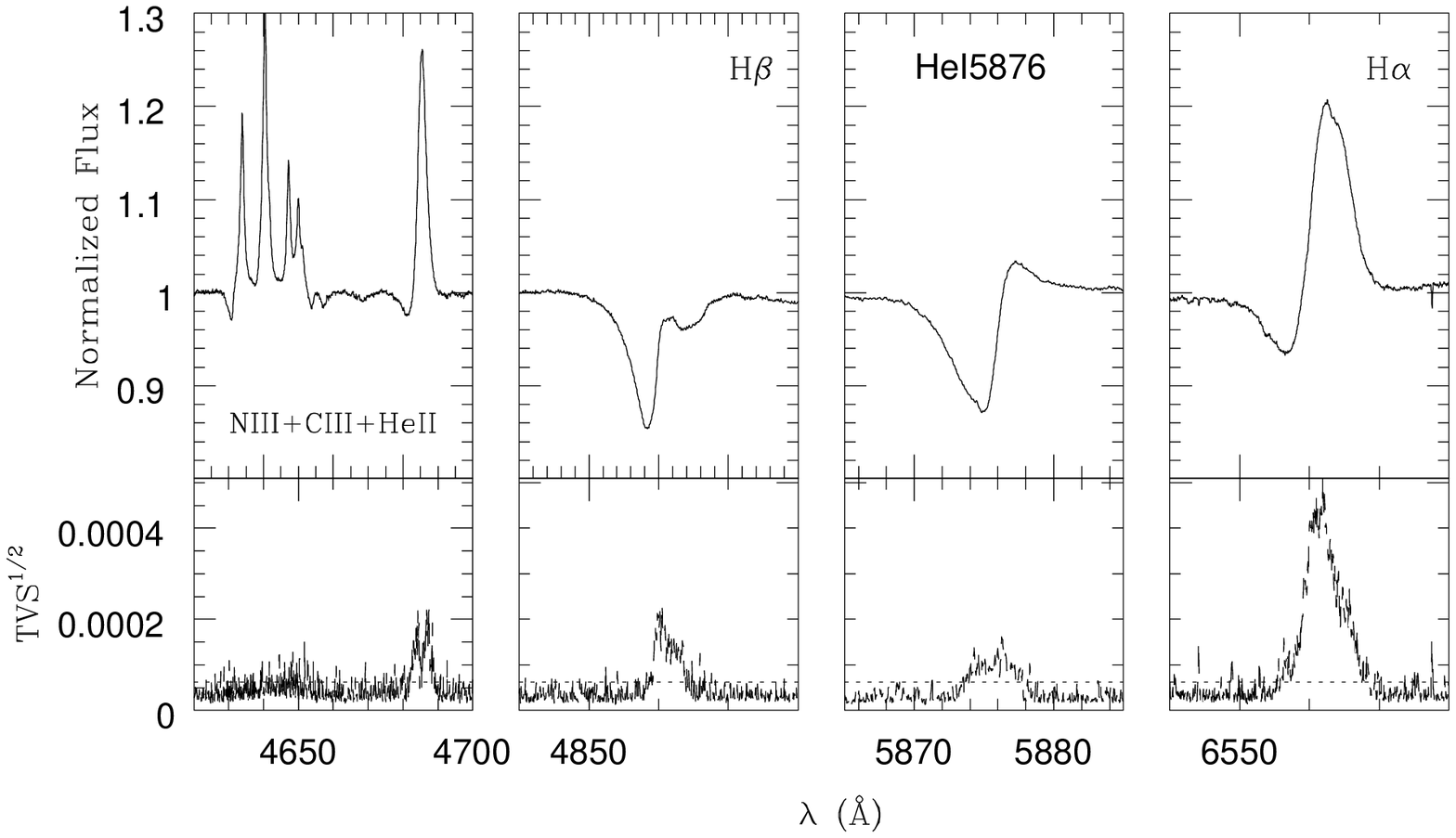}
\includegraphics[width=7.5cm]{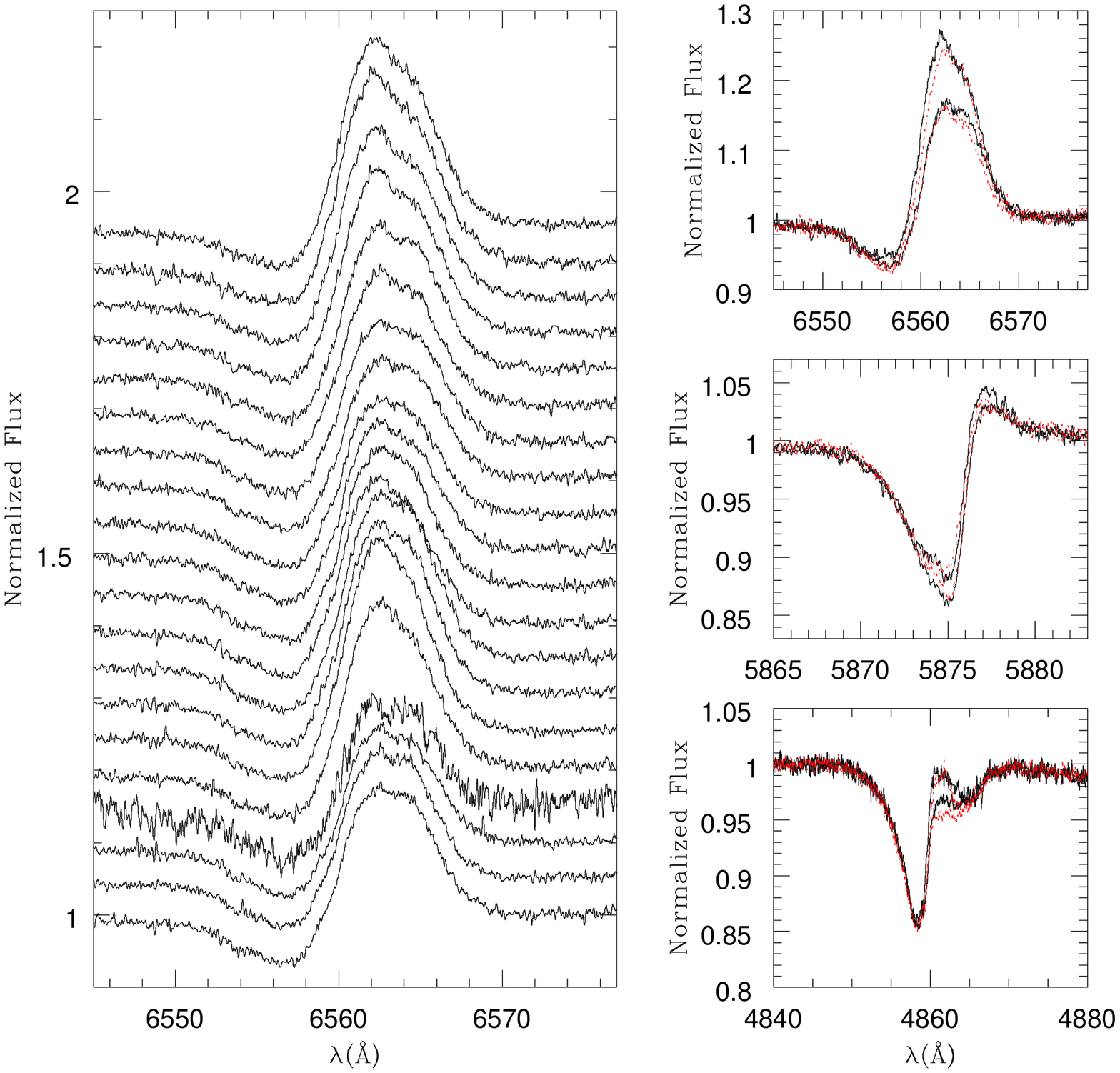}
\caption{\label{148937tvs} Left: Average spectra and TVS of \heii\,$\lambda$\,4686, \hb, \hei\,$\lambda$\,5876, and \ha\ for HD\,148937. The \hei\,$\lambda$\,5876 and \ha\ data were corrected for the contamination by telluric lines. The horizontal dotted line in the lower panels represents the 99\% confidence level. Right: The 20 Coralie observations of the \ha\ line (from May 8 to May 21, note the high noise in the data of May 19, which reflects in the EW measurements) and the extreme states of \ha, \hei\,$\lambda$\,5876, and \hb. The black solid lines correspond to the spectra of May 09 (max emission state) and May 12 (min), the dotted red lines to those of May 15 (max) and May 20 (min). The colored version of the figure is available on-line.} 
\end{center}
\end{figure*}

The variability detected when looking at EWs is confirmed by the Temporal Variance spectrum \citep[TVS, see ][and Fig. \ref{148937tvs}]{ful96}, which compares, in each wavelength bin, the variations between spectra with the scatter expected from random noise. It is important to note that the variability has not the same amplitude for all lines of a given element: for the Balmer lines, the TVS is very significant for \ha, significant for \hb, and barely significant for \hg. This might explain why the variability of, e.g., \hei\,$\lambda$\,4471 is not detected - its amplitude, expected to be lower than for \hei\,$\lambda$\,5876, is certainly hidden by the noise. Note that no significant variation of the \ciii\,$\lambda$\,4650 lines is detected.

\begin{figure*}
\begin{center}
\includegraphics[width=9.2cm]{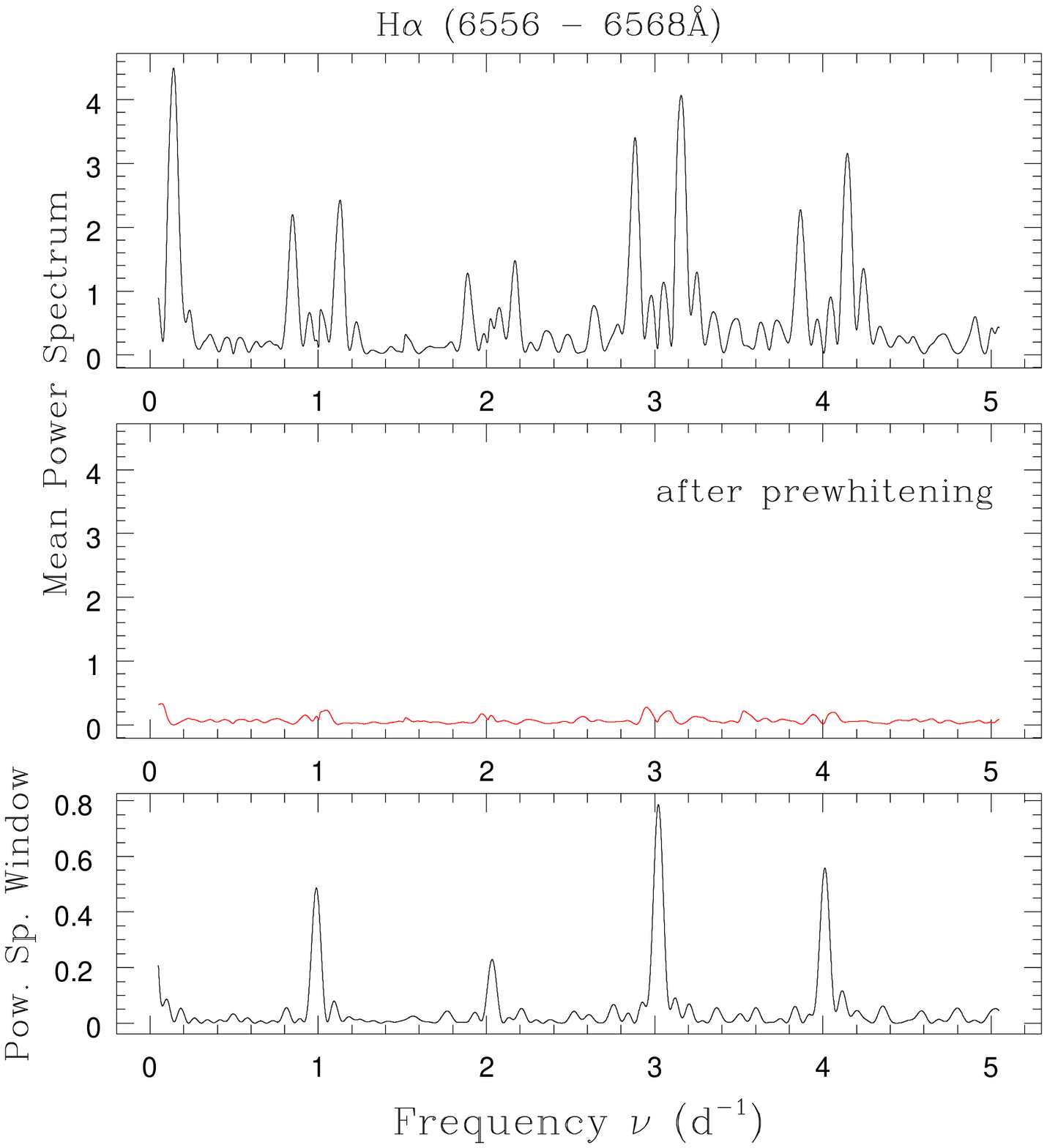}
\includegraphics[width=9cm]{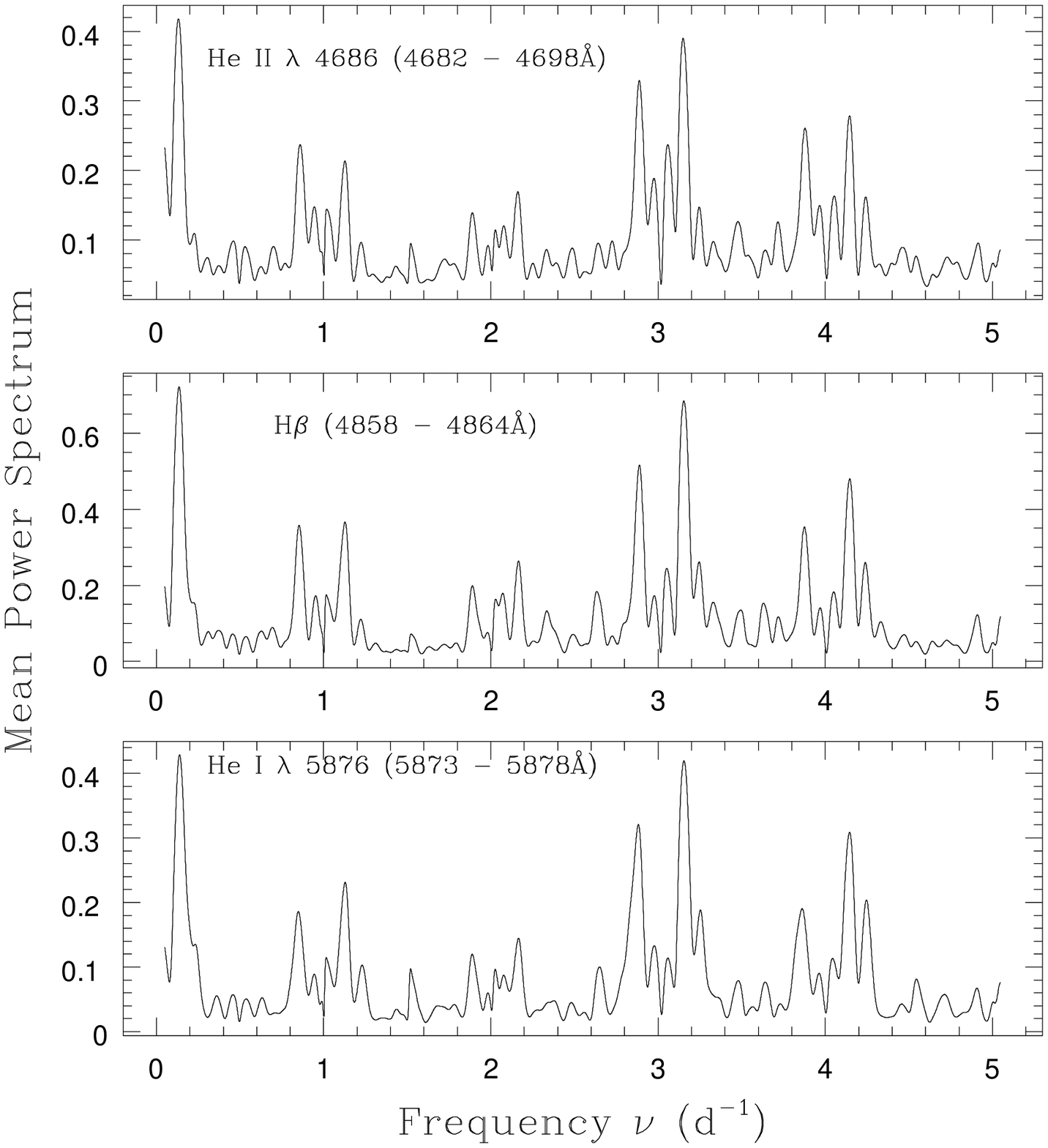}
\caption{\label{148937four} Left, from top to bottom: Raw periodogram of \ha\ as observed during the May 2008 campaign, periodogram after prewhitening by the period of 7.16d, and associated spectral window in the case of HD\,148937. The latter is directly related to the sampling of the time series. Right, from top to bottom: Periodograms of \heii\,$\lambda$\,4686, \hb, and \hei\,$\lambda$\,5876. The power spectra were calculated and averaged in the interval of significant variability (see Fig. \ref{148937tvs} and top label of each panel). The \hei\,$\lambda$\,5876 and \ha\ data were corrected for the contamination by telluric lines.} 
\end{center}
\end{figure*}

Periodograms were calculated for the 20-exposures dataset using the techniques of \citet[see remarks in \citealt{gos01}]{hec85} in the wavelength interval where lines vary significantly \citep[for details on the method see][]{rau01}. The results are shown in Fig. \ref{148937four}: the highest peak is located at a frequency of 0.1315d$^{-1}$, 0.1353d$^{-1}$, 0.1373d$^{-1}$, and 0.1397d$^{-1}$ for \heii\,$\lambda$\,4686, \hb, \hei\,$\lambda$\,5876, and \ha, respectively. These frequencies translate into a period of 7.16--7.60$\pm$0.40d. The relatively large error, compared to the old dataset simply stems from the different observational timespans (2 weeks vs 2.5\,yrs) while the period differences from one line to the other have two origins: the noise, which affects more the fainter lines such as \hei\,$\lambda$\,5876, and the different amplitudes of the variations (the signal is indeed more difficult to catch in case of low amplitude changes, e.g. for \hei\,$\lambda$\,5876). The \ha\ line, which is both strong and the most variable line, yields the most reliable period, i.e. 7.16$\pm$0.40d from the sole new dataset. Taking the errors into account\footnote{Note that combining the two datasets is not possible due to the very different spectral resolutions ($R$=2300 vs 55000). As the best period estimate comes from the longest time coverage, the value of 7.031$\pm$0.003d from \citet{naz08} should be considered as the best measurement on an absolute scale. }, this agrees with the results reported by \citet{naz08}. The 7d period was thus not an artefact from our inadequate time sampling. HD\,148937 therefore appears quite similar to the two other Galactic Of?p stars, though with a shorter period, a smaller amplitude of the variations and with the sole exception of the apparent constancy in the \ciii\ lines. A spectropolarimetric monitoring should be undertaken to see if this could be related to e.g. a low inclination of the star's rotation axis.

\begin{table}
\begin{center}
\caption{ Mean and dispersion of the measured RVs and EWs for HD\,148937. \label{tab:148937}}
\begin{tabular}{lrr}
\hline\hline
Line & RV  & EW \\
& (\kms) & (\AA) \\
\hline
\hg                   &                 &   1.467$\pm$0.044 \\
\heii\,$\lambda$\,4686& $-$14.4$\pm$2.7 &$-$0.697$\pm$0.064 \\
\hb                   &                 &   0.931$\pm$0.061 \\
\civ\,$\lambda$\,5812 & $-$33.6$\pm$2.3 &   0.231$\pm$0.013 \\
\hei\,$\lambda$\,5876 &                 &   0.416$\pm$0.038 \\
\ha                   &                 &$-$0.871$\pm$0.251 \\
DIB\,$\lambda$\,6613  & $-$3.3$\pm$0.5  &   0.143$\pm$0.010 \\
\hline
\end{tabular}
\end{center}
\end{table}

\subsection{HD\,191612}

\subsubsection{X-ray data}

\begin{figure*}
\begin{center}
\includegraphics[width=9cm]{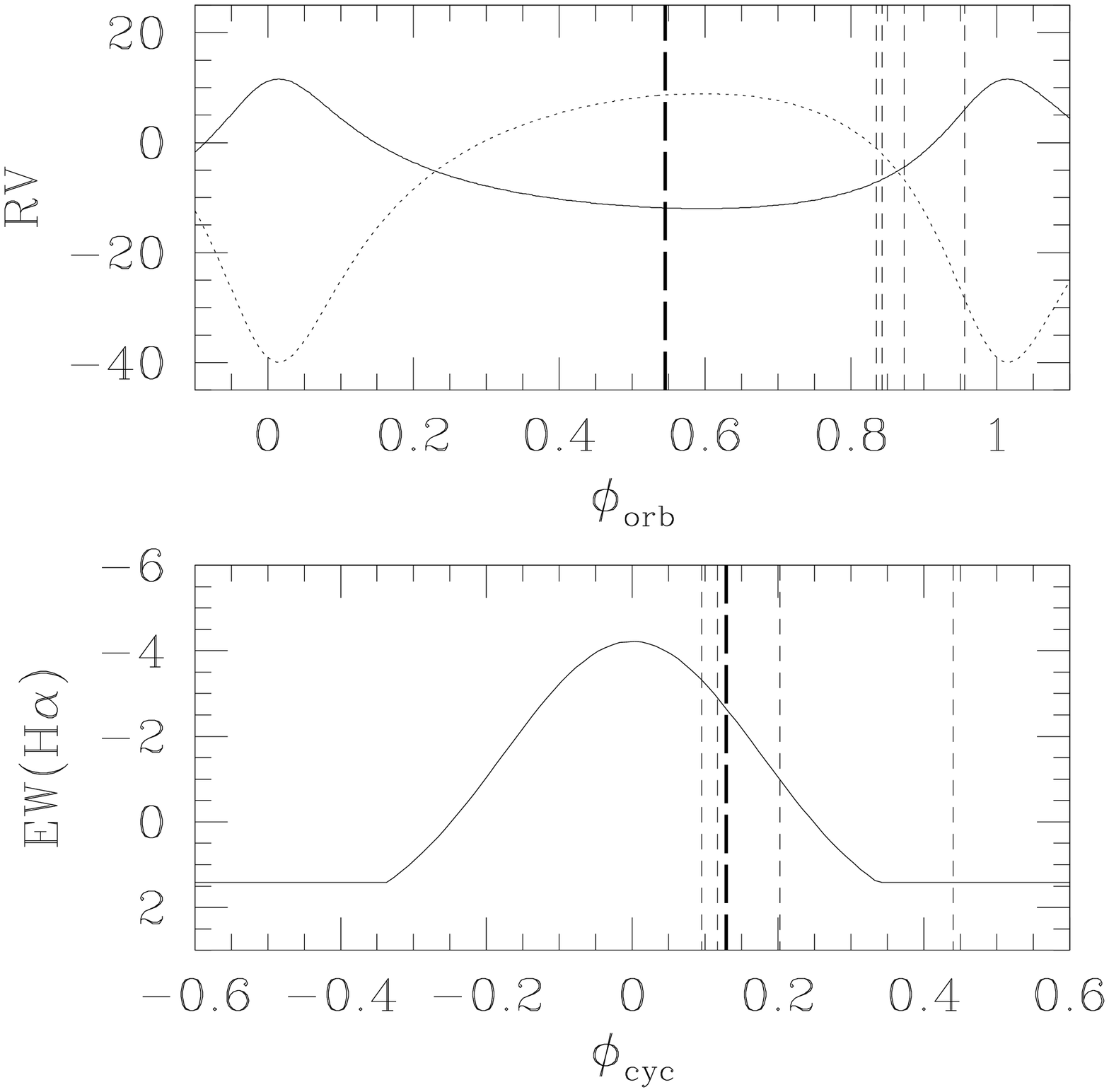}
\includegraphics[width=9cm]{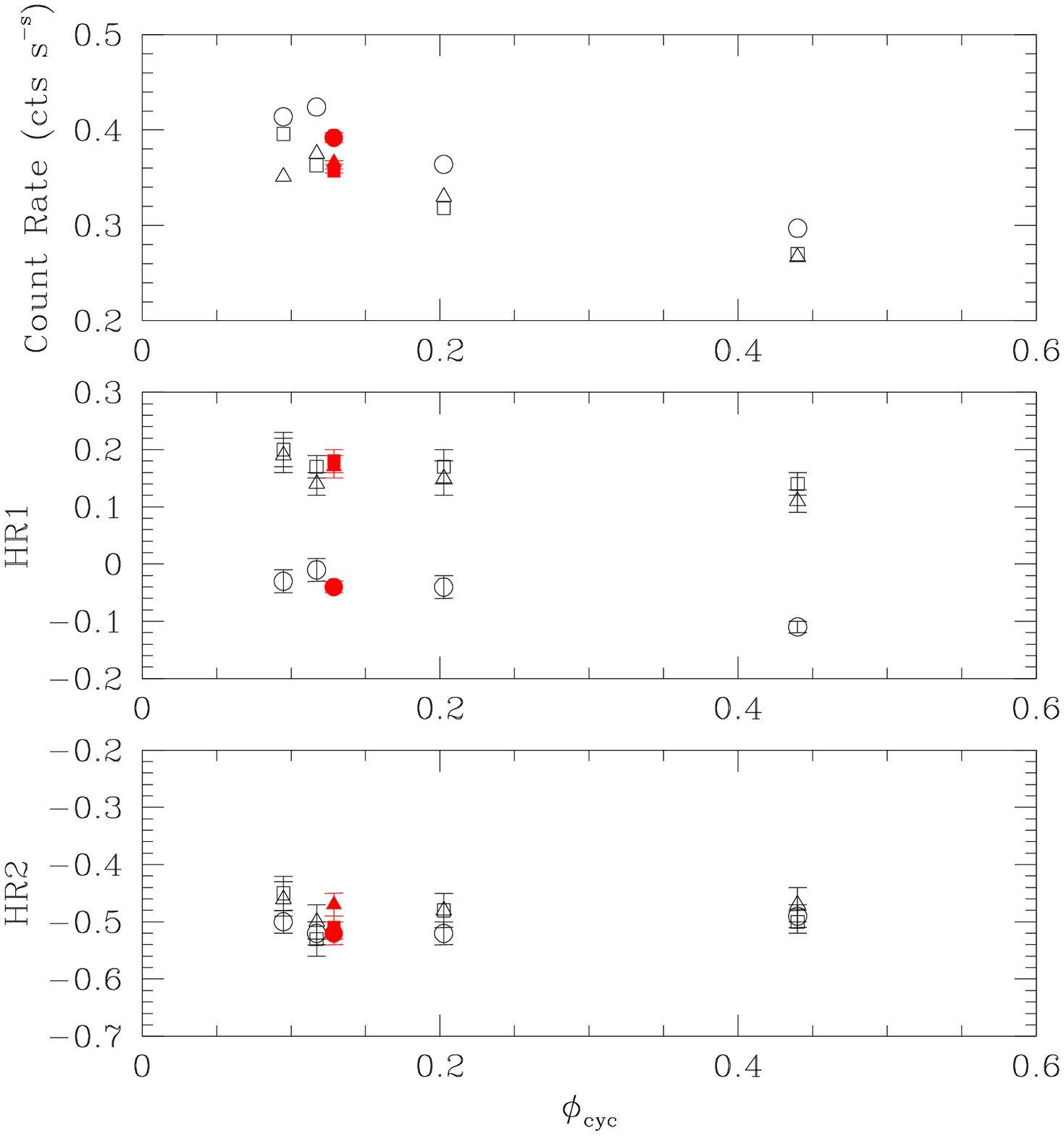}
\caption{\label{191612} Left: Position of the new \xmm\ observation in the cycles of HD\,191612. The orbital cycle, calculated using orbital parameters of \citet{how07} and with the primary (resp. secondary) curve drawn in solid (resp. dotted) line, is shown on top while the bottom panel shows the line profile variation cycle of 538d, from the analytical approximation of \citet{how07}. The phases of the previous observations are shown by the thin short-dashed lines and the phase of the new observation by the thick long-dashed line. Right: Evolution of the count rate in the 0.4--10.0\,keV band (top) and of two hardness ratios (middle, bottom) with phase from the 538d cycle. Open symbols refer to the old observations, filled red symbols to the new dataset; triangles and squares were used for the EPIC MOS data (note that the MOS count rates are multiplied by a factor of 3 in this figure) and circles for the EPIC pn data. The hardness ratio $HR_1$ and $HR_2$ are defined as $(M-S)/(M+S)$ and $(H-M)/(H+M)$, respectively, where $S$ is the count rate in the 0.4--1.0\,keV band, $M$ in 1--2\,keV, and $H$ in 2--10\,keV. The colored version of the figure is available on-line.} 
\end{center}
\end{figure*}

The previous observations of HD\,191612 \citep{naz07} revealed changes in the X-ray flux but the origin of this variability was unclear. To get a better understanding of this phenomenon, we undertook a new \xmm\ observation, whose scheduling was carefully chosen. While the previous observations were taken at orbital phases $\phi_{orb}$=0.83--0.96 and at phases $\phi_{cyc}$=0.09--0.44 in the 538d cycle, the new exposure corresponds to a phase $\phi_{cyc}$=0.13 in the line-profile/photometric cycle of 538d and a phase $\phi_{orb}$=0.55 in the orbital cycle (Fig. \ref{191612}, ephemeris taken from \citealt{how07}). The new data thus enable us to check whether the flux variations are linked to the circumstellar, possibly magnetic, structures surrounding the star or to colliding winds in a binary. In the former case, the new data should be nearly identical to the previous observation taken at a similar phase $\phi_{cyc}$. On the contrary, in the latter case, a dramatic variation is expected. In a colliding wind binary, variations of the flux, correlated with orbital phase, are expected to arise from changes in the separation (in eccentric systems) and/or in the absorbing column (when the winds are different). Indeed, such variations have been observed in several systems \citep[for a review, see][]{gud09}. In HD\,191612, the previous observations, taken before periastron, revealed a 20\% change in reddening-corrected flux (see Table \ref{tab:fit}) for a phase difference $\Delta \phi_{orb}$ of only 0.12, corresponding to a change in separation of 36\%. The new observation was taken two-thirds of the orbit later, near apastron: the change in separation compared to previous data has thus increased (up to 60\%), and so should the difference in flux, if truly linked to the colliding wind emission.

The new data were processed following the same procedure as for the previous data, but using more recent \xmm\ reduction software. The count rates derived from a run of the task $edetect\_chain$ amount to 0.122$\pm$0.002\,cts\,s$^{-1}$ for EPIC MOS1, 0.119$\pm$0.002\,cts\,s$^{-1}$ for EPIC MOS2, and 0.392$\pm$0.005\,cts\,s$^{-1}$ for EPIC pn. The evolution of the count rates and hardness ratio is shown in Fig. \ref{191612}. The new data agree very well with previous observations taken at the same phase of the 538\,d cycle, both in strength (similar count rates) and in hardness (similar hardness ratios) ; they are in addition clearly different from the results of previous observations obtained at the latest phase (where the count rates are $\sim$35\% smaller, corresponding to a $\sim$20$\sigma$ difference, and the hardness ratio $HR_1$ measured on pn data is lower by 7$\sigma$).

The best spectral fit is further provided in Table \ref{tab:fit}, together with the previous results for comparison (reproduced from \citealt{naz07}). The fitted model has the form {\tt wabs($N_{\rm int}^{\rm H}$)*[wabs($N^{\rm H}_1$)*mekal(k$T_1$)+wabs($N^{\rm H}_2$)*mekal(k$T_2$)]}, with {\tt wabs($N_{\rm int}^{\rm H}$)}$=3.4\times10^{21}$~cm$^{-2}$ \citep{dip94} and using solar abundances for both absorptions and optically-thin plasma emissions - i.e. the same model as used in \citet{naz07} for homogeneity reasons\footnote{As in \citet{naz07}, we noted the two sets of parameters giving a good fit, but we recall that a fit by a differential emission measure (DEM) model clearly favors the `cool' fit.}. Quoted fluxes are in the 0.4$-$10.0\,keV energy range, and the unabsorbed fluxes $f_{\rm X}^{\rm unabs}$ are corrected only for the interstellar absorbing column. For each parameter, the lower and upper limits of the 90\% confidence interval (derived from the {\sc error} command under XSPEC) are noted as indices and exponents, respectively. The normalisation factors are defined as $\frac{10^{-14}}{4\pi D^2} \int n_e n_{\rm H} dV$, where $D$, $n_e$ and $n_{\rm H}$ are respectively the distance to the source, the electron and proton density of the emitting plasma. The phases $\phi_{orb}$ and $\phi_{cyc}$ correspond to the phases in the orbital cycle and the 538\,d line profile/photometric cycle, respectively (ephemeris of \citealt{how07}). Again, the spectral fits to the 2008 data agree very well with the previous observations taken at the same phase of the 538\,d cycle, yielding the same X-ray luminosity, 9$\times10^{32}$\,erg\,s$^{-1}$ (for a distance of 2.29\,kpc), and same \loglxlb, $-$6.1.

The conclusions are obvious: despite the very different orbital phases, the new data are nearly identical to those taken 3 years before at a similar phase in the 538d cycle. The modulation of the X-ray emission thus appears directly correlated with the optical line-profile variations, a situation very similar to that observed for $\theta^1$\,Ori\,C \citep{gag05,naz08b}. 

\begin{table*}
\centering
\caption{Best-fitting models and X-ray fluxes at Earth for each \xmm\ observation of HD\,191612. 
%The fitted model has the form {\tt wabs($N_{\rm int}^{\rm H}$)*[wabs($N^{\rm H}_1$)*mekal(k$T_1$)+wabs($N^{\rm H}_2$)*mekal(k$T_2$)]}, with {\tt wabs($N_{\rm int}^{\rm H}$)}$=3.4\times10^{21}$~cm$^{-2}$ \citep{dip94}. Quoted fluxes are in the 0.4$-$10.0\,keV energy range, and the unabsorbed fluxes $f_{\rm X}^{\rm unabs}$ are corrected only for the interstellar absorbing column. For each parameter, the lower and upper limits of the 90\% confidence interval (derived from the {\sc error} command under XSPEC) are noted as indices and exponents, respectively. The normalisation factors are defined as $\frac{10^{-14}}{4\pi D^2} \int n_e n_{\rm H} dV$, where $D$, $n_e$ and $n_{\rm H}$ are respectively the distance to the source, the electron and proton density of the emitting plasma. The phases $\phi_{orb}$ and $\phi_{cyc}$ correspond to the phases in the orbital cycle and the 538\,d line profile/photometric cycle, respectively (ephemeris of \citealt{how07}). The 2005 results are reproduced from \citet{naz07}.
\label{tab:fit}}
\setlength{\tabcolsep}{1.5mm}
\begin{center}
\begin{tabular}{cccccccccclcc}
            \hline\hline
Date & Rev. & $\phi_{cyc}$ & $\phi_{orb}$ & $N_1^{\rm H}$  & k$T_1$   & $norm_1$     & $N_2^{\rm H}$  & k$T_2$ & $norm_2$     & $\chi^2_{\rm \nu}$(dof) & $f_{\rm X}^{\rm abs}$ & $f_{\rm X}^{\rm unabs}$\\
&     & & & $10^{22}$~cm$^{-2}$ & keV   & $10^{-3}$cm$^{-5}$         & $10^{22}$~cm$^{-2}$ & keV  & $10^{-3}$cm$^{-5}$        &                               & \multicolumn{2}{c}{($10^{-13}$~erg\,cm$^{-2}$\,s$^{-1}$)}  \\
\hline
\multicolumn{10}{l}{`cool' model}\\
\vspace*{-0.3cm}&&&&&&\\
05/04/05 & 975  & 0.09 & 0.84 & $0.51_{0.42}^{0.61}$ & $0.23_{0.20}^{0.25}$ & $7.07_{4.04}^{15.9}$ &  $1.07_{0.87}^{1.31}$ & $1.27_{1.19}^{1.36}$ & $1.08_{0.96}^{1.21}$ & 1.16 (411) & 7.0 & 14.1\\
\vspace*{-0.3cm}&&&&&&\\
17/04/05 & 981  & 0.12 & 0.84 & $0.32_{0.25}^{0.46}$ & $0.27_{0.23}^{0.29}$ & $2.15_{1.40}^{3.81}$ & $0.91_{0.80}^{1.10}$ & $1.01_{0.96}^{1.07}$ & $1.22_{1.11}^{1.35}$ & 1.10 (681) & 6.8 & 14.3\\
\vspace*{-0.3cm}&&&&&&\\
02/06/05 & 1004 & 0.20 & 0.87 & $0.43_{0.31}^{0.51}$ & $0.25_{0.23}^{0.28}$ & $3.78_{1.78}^{5.83}$ & $1.03_{0.81}^{1.25}$ & $1.22_{0.98}^{1.31}$ & $0.89_{0.74}^{1.11}$ & 1.15 (460) & 6.0 & 12.5\\
\vspace*{-0.3cm}&&&&&&\\
08/10/05 & 1068 & 0.44 & 0.96 & $0.61_{0.58}^{0.64}$ & $0.20_{0.19}^{0.21}$ & $13.6_{10.7}^{18.1}$ & $1.39_{1.14}^{1.68}$ & $1.60_{1.49}^{1.72}$ & $0.66_{0.59}^{0.76}$ & 1.31 (555) & 5.3 & 11.0\\
\vspace*{-0.3cm}&&&&&&\\
03/04/08 & 1523 & 0.13 & 0.55 & $0.51_{0.46}^{0.56}$ & $0.24_{0.23}^{0.25}$ & $6.67_{4.75}^{9.22}$ & $1.21_{1.05}^{1.41}$ & $1.34_{1.28}^{1.41}$ & $1.06_{0.97}^{1.17}$ & 1.31 (591) & 7.2 & 14.4\\
\vspace*{-0.3cm}&&&&&&\\
\hline
\multicolumn{10}{l}{`hot' model}\\
\vspace*{-0.3cm}&&&&&&\\
05/04/05 & 975 & 0.09 & 0.84 & $0._{0.}^{0.05}$ & 0.64$_{0.62}^{0.66}$ & 0.28$_{0.27}^{0.33}$ & $0._{0.}^{0.03}$ & $2.35_{2.21}^{2.54}$ & $0.51_{0.48}^{0.53}$ & 1.19 (411) & 7.6 & 14.4\\
\vspace*{-0.3cm}&&&&&&\\
17/04/05 & 981  & 0.12 & 0.84 & $0._{0.}^{0.01}$ & $0.60_{0.58}^{0.61}$ & $0.32_{0.31}^{0.34}$ & $0.81_{0.70}^{0.87}$ & $1.23_{1.06}^{1.32}$ & $0.95_{0.86}^{1.15}$ & 1.08 (681) & 7.0 & 13.9\\
\vspace*{-0.3cm}&&&&&&\\
02/06/05 & 1004 & 0.20 & 0.87 & $0._{0.}^{0.01}$ & $0.51_{0.47}^{0.54}$ & $0.28_{0.26}^{0.30}$ & $0.82_{0.74}^{0.91}$ & $1.07_{1.01}^{1.27}$ & $0.92_{0.75}^{1.01}$ & 1.16 (460) & 5.8 & 12.0\\
\vspace*{-0.3cm}&&&&&&\\
08/10/05 & 1068 & 0.44 & 0.96 & $0._{0.}^{0.01}$ & $0.61_{0.60}^{0.63}$ & $0.23_{0.22}^{0.24}$ & $0._{0.}^{0.01}$ & $2.53_{2.38}^{2.68}$ & $0.32_{0.31}^{0.34}$ & 1.39 (555) & 5.4 & 11.0\\
\vspace*{-0.3cm}&&&&&&\\
03/04/08 & 1523 & 0.13 & 0.55 & $0._{0.}^{0.01}$ & $0.61_{0.60}^{0.62}$ & $0.33_{0.32}^{0.35}$ & $0.57_{0.44}^{0.69}$ & $1.64_{1.53}^{1.76}$ & $0.73_{0.66}^{0.79}$ & 1.31 (591) & 7.3 & 14.1\\
\vspace*{-0.3cm}&&&&&&\\
\hline
\end{tabular}
\end{center}
\end{table*}

\subsubsection{MHD modelling}

To better understand the properties of HD\,191612, a fully-dynamic numerical magneto-hydrodynamic (MHD) model, taking into account the impact on the wind of the magnetic field detection, is needed. We used the publically available ZEUS-3D code \citep{sto92} to evolve a consistent dynamical solution for a line-driven stellar wind from the star with a dipolar surface field. The stellar parameters of HD\,191612, adopted mostly from \citet[and references therein]{naz08b}, are listed in Table \ref{tab:param}.

HD\,191612 is a rather slow rotator with a surface rotation speed $<$50\kms\ \citep{how07,naz08b}. Such a slow rotation is unlikely to have any significant dynamical effects on the stellar wind. Therefore, the wind was assumed to be azimuthally symmetric with the magnetic field axis aligned with the rotation axis\footnote{This simplifies the computation and makes no assumption on the true geometry of the system, i.e. an oblique rotator model is not excluded ; it can easily be simulated here by changing the viewing angle on the system.}. This is essentially  a '2.5-D' formulation allowing for non-zero azimuthal components of both the magnetic field $B_{\phi}$ and velocity $v_{\phi}$, while still assuming that all quantities are constant in the azimuthal coordinate angle $\phi$. 

Much of the numerical procedures in this simulation follows \citet{udd08}. The models thus include rotational effects but, for the first time, we also use the full energy equation, with adiabatic index $\gamma$=5/3 and radiative cooling of \citet[][which assumed solar abundance]{mac81}\footnote{Note that models used by \citet{gag05} made use of the full energy equation but did not include rotation.}. At the initial time $t=0$, the wind is represented by a relaxed non-magnetic and non-rotating model, but we then simultaneously introduce a dipole magnetic field and a surface rotation at the lower boundary, both defined relative to a common polar axis. The wind and magnetic field are then let free to compete with each other. To ensure that there are no numerical artefacts arising from the initial conditions, we run our model for about 1.5\,Msec, which is at least 50 fluid crossing times.

Immediately after $t=0$, the magnetic field channels wind material towards the tops of closed loops near the equator. There, the collision with the opposite stream leads to strong shocks, like in ``magnetically confined wind shocks'' (MCWS) presented by \citet{bab97}. Denser material then cools radiatively very quickly leading to a dense disk-like structure (Fig. \ref{snapshot}). Some of the less dense post-shock material within the closed magetic loops remain very hot (ca. 80MK) for a long period of time. In this model, the Alfv\'en radius or magnetosphere extends about 3--4 stellar radii above the stellar surface.

Initially mass builds up in the equatorial region below the Keplerian radius estimated to be at about 4.5\,$R_*$, but then there appear repeated episodes of infall of inner disk material back onto the stellar surface. Over a somewhat longer timescale, there appears another, somewhat different kind of disruption, one that starts higher up, closer to the Alfv\'en radius. This is characterized by outward ejections of the upper disk mass, leading to `centrifugally driven mass ejections'. Such ejections can lead to X-ray flares (see \citealt{udd06}) but these short-term variations cannot be detected in the observations of HD\,191612, since they lack the required sensitivity to analyze the lightcurve on very short timescales.

\begin{figure*}
\begin{center}
\caption{\label{snapshot} Snapshots of density (left) and temperature (right), in cgs units on a logarithmic color-scale, shown with field lines (solid lines) at arbitrarily chosen time t=1300\,ks for the 2D MHD simulation of HD\,191612. the x and y axes are labelled in units of the stellar radius. Note the unsteady nature of the dense equatorial region which undergoes episodic emptyings and refills. Within the magnetosphere (i.e. the closed loop region), the opacity at 10\AA\ or 1.2\,keV may reach high values ($\tau$=10 for $\kappa$=30\,cm$^2$\,g$^{-1}$, see \citealt{coh10}, a density of 10$^{-12}$\,g\,cm$^{-3}$ and a disk height of 0.03\,$R_*$) but it is totally negligible in the dense equatorial region outside magnetosphere. The observations show only a small absorption in addition to the interstellar extinction and it is not significantly varying with phase (i.e. there is no clear evidence of excess absorption in the magnetic equatorial plane, or the effect is much smaller than would be expected from a dense equatorial cooling disk, as found in \citealt{gag05}). } 
\end{center}
\end{figure*}

\begin{figure*}
\begin{center}
\caption{\label{sim} Volume emission-measure distribution (DEM) per log T= 0.1 bin for our MHD simulations of HD\,191612. The MHD simulation is used to compute the emission measure per unit volume per logarithmic temperature bin, then integrated over three-dimensional space assuming azimuthal symmetry about the magnetic dipole axis. Left: DEM as a function of time and temperature ; the periodic variations due to build-ups and infalls are clearly seen. The colored version of this figure is available on-line. Right: Time-averaged DEM as a function of temperature. } 
\end{center}
\end{figure*}

Fig. \ref{sim} shows the differential emission measure (DEM, or the relative amount of material in a given temperature bin) calculated in the MHD simulation.  The DEM peaks at very high temperatures (about 10\,MK and larger), suggesting that the very hot gas dominates the X-ray emission of the system. Of course, the material in the unshocked regions remain at the effective temperature of the star due to the intense UV radiation from the parent star that keeps the wind nearly isothermal.

\begin{table}
\centering
\caption{Stellar parameters used for the MHD simulation of HD\,191612, adopted mostly from \citet[and references therein]{naz08b}.
\label{tab:param}}
\setlength{\tabcolsep}{1.5mm}
\begin{center}
\begin{tabular}{lr}
            \hline\hline
Parameter & Value\\
\hline
$R$     & 1.015$\times 10^{12}$cm = 14.4\,R$_{\odot}$ \\
$M$     & 7.0$\times 10^{34}$g = 35\,M$_{\odot}$\\
$L$ & 7.697$\times 10^{38}$erg\,s$^{-1}$ = 2$\times 10^5$\,L$_{\odot}$\\
$\alpha$  & 0.55\\
$\bar{Q}$    & 2000\\
$\delta$  & 0.0\\
$\dot M$ & 1$\times 10^{-6}$\,M$_{\odot}$\,yr$^{-1}$\\
$v_{\rm rot}$   & 50\kms\\
$B_{\rm polar}$     & 1500\,G\\
$\eta_*$      & 31 $\sim \sqrt{1000}$\\
\hline
\end{tabular}
\end{center}
\end{table}

\subsubsection{Comparison of models and observations}

Applying a magnetically-confined wind model to HD\,191612 yields four clues. First, the X-ray emission should clearly be dominated by very hot plasma, as was the case for $\theta^1$\,Ori\,C \citep[see Fig. 6 of ][]{gag05}. Second, on average, the total EM calculated for HD191612 in the 1-100MK range amounts to 5.3$\times10^{56}$\,cm$^{-3}$. This agrees well with the MHD simulation of $\theta^1$\,Ori\,C, considering the difference in the stellar parameters and modelling. Indeed, the simulated EM for this star, in the same temperature range, reached 8.2$\times10^{55}$\,cm$^{-3}$ \citep{gag05}, i.e. a factor 6.5 lower than for HD\,191612 which is close to the factor of 4 expected from the EMs scaling with the third power of the radius. Considering an average cooling factor $\Lambda(T)$ of $\sim2\times10^{-23}$\,erg\,cm$^3$\,s$^{-1}$ \citep{ray76}, this yields a {\it totally unabsorbed} X-ray luminosity of $\sim8\times10^{33}$\,erg\,s$^{-1}$ in the 0.5--10.\,keV range. Third, in the MHD models, the hot gas generally moves with velocities of about 100--200\kms\ and only very few hot plasma reach 400\kms, whatever the viewing angle on the equatorially-confined region. This is unsurprising since the hot gas always remains close to the star: the associated X-ray lines are thus expected to be narrow, again a similar result as that found for $\theta^1$\,Ori\,C. Finally, the X-ray emission should be modulated throughout the cycle of 538d, as the viewing angle on the disk-like structure changes.

The observations (both old and new) do not fully agree with this picture. On the one hand, the emission is modulated with the 538d cycle, as expected for MCWS. The observed X-ray emission is indeed bright ($L_X$=7--9$\times10^{32}$\,erg\,s$^{-1}$ when corrected for the interstellar absorption), in fact on the luminous edge compared to ``normal'' O-type stars (for O stars observed with \xmm, \loglxlb =$-6.45$ with a dispersion of 0.51 dex, see \citealt{naz09}). Considering the uncertainty in the spectral models (see Table \ref{tab:fit}), the observations yield a {\it totally unabsorbed} X-ray luminosity of $\sim$8--80$\times10^{32}$\,erg\,s$^{-1}$, for the `hot' and `cool' model respectively. Total emission measures, derived from the normalization factors of Table \ref{tab:fit} or as in \citet{gag05} for a spectral fitting considering a single absorbing column for both thermal components, amount to 3--7$\times10^{56}$\,cm$^{-3}$ for an equivalent `cool' model and ten times lower for an equivalent `hot' model. Theoretical predictions thus agree rather well with the `cool' model, which is also the one favored by the DEM modelling\citep{naz07}. 

On the other hand, however, the X-ray emission of HD\,191612 displays broad lines \citep[$FWHM_{obs}\sim$2000\kms, ][]{naz07} and is far from being as hard as predicted by the model. In fact, a fit of the EPIC data by a DEM model yields a strong emission peak at about 2\,MK, not beyond 10\,MK, as shown for example in Fig. 3 of \citet{naz07}. In addition, the average temperature\footnote{$\sum kT_i\times norm_i/\sum norm_i$ for models using a single absorption in front of the thermal emission components} derived in the 2XMM survey for HD\,191612 amounts to $\sim$0.3\,keV and is thus quite quite normal (85\% of the stars in the 2XMM have $<kT>$ below 1\,keV, \citealt{naz09}) - it does not resemble the extreme case of  $\theta^1$\,Ori\,C for which $<kT>$ amounts to 2.5\,keV in the same 2XMM survey and for which the MCWS model works well. The relatively soft character of that emission despite the large magnetic confinement parameter might be reminiscent of the characteristics observed in some B-type stars \citep{gud09}: further investigation of the MCWS model and magnetic field geometry is thus required to understand the high-energy properties of the Of?p stars.

\section{Conclusions}
Our continued monitoring in the X-ray and optical domains of the prototypical Of?p stars has now clarified some of their observational properties. 
 
For HD\,108, the Balmer, \ciii\,$\lambda$4650, \hei, and Si\,{\sc iii} lines display minimum emissions since 2005: the star has thus finally reached its quiescent state, for the first time in 50--60yrs. Long-term monitoring of this quiescence will further constrain the geometry of the line emission region.

For HD\,148937, the presence of the 7d variations in the Balmer lines is confirmed by a short-term monitoring. During this cycle, the amplitude of the emission peak in \ha, the most variable line, changes by only 7\%. Thanks to the high spectral resolution of the new data, similar periodic variations, though of even smaller amplitude, are detected in the \hei\,$\lambda$\,5876 and \heii\,$\lambda$\,4686 lines. These detections underline the similarities of the star with the other two prototypical Of?p objects. 

For HD\,191612, a new \xmm\ observation clearly shows that its X-ray emission is modulated by the 538d period and not the orbital period - it is thus not of colliding-wind origin: the phenomenon responsible for the optical changes appears also at work in the high-energy domain. Contrary to observations, our MHD simulations  predict a hard spectrum, dominated by very hot plasma ($>$10\,MK) and narrow lines. Further refinements to the modelling will be needed in order to fully explain this discrepancy.

\begin{acknowledgements}
YN, MDB and GR acknowledge support from the Fonds National de la Recherche Scientifique (Belgium), the Communaut\'e Fran\c caise de Belgique, the PRODEX XMM and Integral contracts, and the `Action de Recherche Concert\'ee' (CFWB-Acad\'emie Wallonie Europe). The authors thank L. Mahy and P. Eenens for providing the 2009 \'echelle spectrum. ADS and CDS were used for preparing this document.
\end{acknowledgements}

\end{document}